\documentclass[aps,prb]{revtex4}

\usepackage{graphicx}
\usepackage{epsfig}
\usepackage{dcolumn}
\usepackage{bm}

\begin{document}

\title{Propagating and evanescent waves in absorbing media}
\author{S. Anantha Ramakrishna}
\affiliation{The Blackett Laboratory, Imperial College, London SW7 2BW, United Kingdom}
\author{A. D. Armour}
\affiliation{School of Physics and
Astronomy, University of Nottingham, Nottingham NG7 2RD, United Kingdom}


\begin{abstract}
We compare the behavior of propagating and evanescent light waves in absorbing media
with that of electrons in the presence of inelastic scattering. The imaginary part of
the dielectric constant results primarily in an exponential decay of a propagating
wave, but a phase shift for an evanescent wave. We then describe how the scattering of
quantum particles out of a particular coherent channel can be modeled by introducing
an imaginary part to the potential in analogy with the optical case. The imaginary
part of the potential causes additional scattering which can dominate and actually
prevent absorption of the wave for large enough values of the imaginary part. We also
discuss the problem of maximizing the absorption of a wave and point out that the
existence of a bound state greatly aids absorption. We illustrate this point by
considering the absorption of light at the surface of a metal.
\end{abstract}

\maketitle

\section{Introduction}
Analogies between the propagation of classical electromagnetic
waves and the non-relativistic quantum mechanical motion of matter
have been exploited for many years.\cite{schiff} The analogies
stem from the fact that the wave equations for light and quantum
particles have a very similar form.

If the propagation of an electromagnetic wave preserves the
polarization, then Maxwell's equations in the steady state
reduce to the Helmholtz equation for a scalar wave,
\begin{equation}
\nabla^2 {\cal E}({\bf{r}}) + (\omega/c)^2[n^2({\bf{r}})]{\cal
E}({\bf{r}}) = 0,
\end{equation}
where $\cal{E}$ is the complex wave amplitude and $n({\bf{r}})$ is
the refractive index of the medium. If we replace $\cal{E}$ by
$\Psi$, the wavefunction for an electron, $k$ by
$\sqrt{2mE/\hbar^2}$, and $n^2$ by $1-V/E$, where $V$ is the
potential and $E$ is the energy of the electron, we obtain the
time independent Schr\"{o}dinger equation for an electron.  Hence
we see that the effects of the refractive index on electromagnetic
waves and the effects of an electrostatic potential on electrons
are similar.

Despite the close similarity between the equations describing the propagation of
electromagnetic waves and electrons, there is by no means an exact mapping of
properties between the two systems. A particularly tricky topic is the absorption of
waves that takes place in the presence of dissipation, particularly for evanescent
waves. Continuum electromagnetic theory adds an imaginary part to the dielectric
permittivity or the magnetic permeability to model the microscopic processes of
absorption.\cite{foot} An imaginary potential is frequently adopted to model
scattering of electrons into inelastic channels\cite{schiff} and is often referred to
as an optical potential because of the analogy with electromagnetic waves. Such
potentials are extensively used in nuclear physics to describe the scattering events
in which particles are removed from the incident flux and to describe the formation of
complex nuclei.\cite{hodgson}

It is now possible to fabricate solid-state structures that are
small enough for the transport of electrons to be largely coherent
so that the wave nature of the electrons dominates. Such systems
are known as mesoscopic and their current characteristics are
modeled using a scattering formalism analogous to that used to
calculate the transmission of light through one or more slabs of
dielectric material.\cite{datta} The propagation of electrons can
be thought of as due to contributions from a number of different
coherent channels that are essentially coherent wave modes.
Complex potentials can be used in such systems to describe the
removal of electrons from a coherent channel due to inelastic
scattering processes\cite{stonelee} and dephasing\cite{zohta}
(scattering out of one wave mode into others and loss of
coherence). However, although it is frequently used, this approach is
over simplistic because it neglects the important fact that any
potential (whether real or imaginary) also gives rise to
additional scattering.\cite{rk}

We will discuss some aspects of wave propagation in
dissipative media that many students (and some teachers) do not
appreciate after a first course on quantum mechanics. In Sec.~II we begin by
discussing the familiar cases of propagating and evanescent classical
electromagnetic waves in an absorbing medium. In Sec.~III, we analyze the
analogous case for electrons: propagation in a region with an imaginary
contribution to the potential. Such non-unitary Hamiltonians can
formally cause the absorption of the particle and are frequently
used to mimic absorption. We examine the consequences of an
optical potential on both propagating and evanescent electron
waves. Then in Sec.~IV we use the example of scattering from a
complex delta-function potential to explore the
additional scattering that is encountered when using optical
potentials in quantum mechanics. In Sec.~V, we address the
problem of how to construct a maximally absorbing material, a
so-called perfect absorber. Intriguingly, a way of maximizing the
absorption of light in non-magnetic materials can be derived by
analogy with a mesoscopic system. Finally we draw conclusions in
Sec.~VI.

\section{The case of light}
It is possible to describe the propagation of light through an
absorbing medium using a very simple
prescription:\cite{bornandwolf} just add an imaginary part to the
refractive index of the medium, $n=n_r+i n_i$, where $n_i$, the
imaginary part of the refractive index, is proportional to the
degree of absorption in the medium. The ratio $\kappa = n_i/n_r$
is often called
the attenuation index.\cite{bornandwolf} A complex refractive index implies that
the wave vector of a wave in the medium will also be complex,
$k\equiv 2\pi n/\lambda=2\pi(n_r+i n_i)/\lambda$, and a propagating
wave, $e^{ikz}$ (in one dimension), decays in
amplitude as it travels through the medium.

The imaginary part of $n$ enters the continuum theory as an
imaginary component of the dielectric constant.\cite{bornandwolf}
A material will typically be at least partly absorbing and so the
dielectric constant will generally be complex, with the degree of
absorption measured by the magnitude of its imaginary component
(for non-magnetic materials.\cite{fn1}) The refractive index is
given by
\begin{eqnarray}
n&=&c\sqrt{\mu\varepsilon}=c\sqrt{\mu(\varepsilon_r+i\varepsilon_i)} \\
&\simeq&c\sqrt{\mu
\varepsilon_r}\bigl[1+i\frac{\varepsilon_i}{2\varepsilon_r}
\bigr]=n_r+i n_i,
\end{eqnarray}
where $n_i=\frac{c}{2}\sqrt{\frac{\mu}{\varepsilon_r}}\varepsilon_i$,
and we have made the usual assumption that
$\varepsilon_i\ll \varepsilon_r$. The energy flow through the
medium is determined by the magnitude of the time-averaged
Poynting vector for the electromagnetic wave, $\langle {\bf
S}\rangle_t$. For a dissipative medium,
\begin{equation}
|\langle {\bf S}\rangle_t|=\frac{c}{8\pi}|{\bf {\cal E}}\times {
\bf {\cal H}}^*|=\frac{c}{8\pi}\sqrt{\frac{\epsilon}{\mu}}|{\cal
E}_0|^2 e^{-4\pi n_i z/\lambda},
\end{equation}
where $z$ is the distance propagated, $|{\cal E}_0|$ is the
amplitude of the electric field vector associated with the
electromagnetic wave and ${\bf{\cal H}}^*$ is the complex
conjugate of ${\bf{\cal H}}$, the magnetic field vector. The decay
of the energy flow with distance corresponds to the energy
dissipated into the medium ($\sim -dS/dz$). The quantity
$\chi=4\pi n_i/\lambda = 4\pi n_r \kappa/\lambda$ is called the
absorption coefficient.\cite{bornandwolf}

\begin{figure}[h]
\includegraphics[height=200pt,width=200pt]{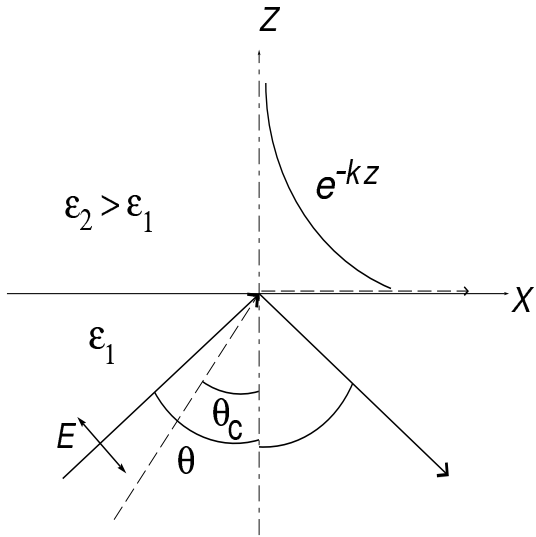}
\caption{\label{fig1}A schematic illustration of total internal
reflection. A wave incident on a dielectric medium with a larger
dielectric constant at an angle greater than the critical angle,
$\theta>\theta_c$, results in an exponentially decaying wave in
medium 2. The plane of incidence is the $x$-$z$ plane and the
electric vector is taken to lie in the plane of polarization
($p$-polarization).}
\end{figure}

Evanescent waves occur when the magnitude of the transverse
incident wave vector, $k_{\parallel}$, is larger than the
wave vector in the medium, $k_{0}=2\pi n/\lambda$. The physics of
evanescent waves is readily understood by considering what happens
when light is shone on a non-absorbing medium at an angle greater
than the angle for total internal reflection from a medium with
lower refractive index. The geometry is illustrated in Fig.~1
where a wave with a wave vector $[k_x^{(1)},0,k_z^{(1)}]$ in medium 1 is incident on
medium 2.
Maxwell's equations require that $(k_x^{(1)})^2
+(k_z^{(1)})^2={\varepsilon_1 \mu}
\bigl(\frac{\omega}{c}\bigr)^2$ in medium 1 and the
continuity condition at the boundary requires $k_x^{(2)}
=k_x^{(1)}=k_x$. Therefore, if $\varepsilon_2<\varepsilon_1$ and
$k_x$ is sufficiently large, $k_z^{(2)}$ must become imaginary to satisfy the
condition in medium 2,
$(k_x^{(2)})^2+(k_z^{(2)})^2={\varepsilon_2 \mu}\bigl(\frac
{\omega}{c}\bigr)^2$.
This condition is the origin of total internal reflection as
there exists no propagating wave inside medium 2 for
$k_x>\sqrt{\varepsilon_2 \mu}\bigl(\frac{\omega}{c}\bigr)$.

The spatial
dependence of the electromagnetic field in medium 2 takes the form
\begin{equation}
e^{i{\bf k}^{(2)}.{\bf r}}=e^{ ik_x^{(2)}x-|k_z^{(2)}|z},
\end{equation}
and the amplitude of the wave decays exponentially with
distance into the medium. However, in this case the
decay of the wave amplitude does not imply the absorption of the wave
energy because examination of the Poynting vector shows that no energy
enters medium 2. Evanescent waves are the classic example of a
wave in which exponential decay of the amplitude does not imply
that dissipation is taking place.

The electromagnetic near-field of a radiation source and the
electromagnetic fields inside a metal or plasma are also
evanescent.\cite{bornandwolf} The case of the field inside a metal
or plasma is interesting because the dielectric constant is large, is
almost entirely real, but with a negative sign. A negative
dielectric constant makes it impossible to satisfy the dispersion
relation, $k^2_x+k^2_y+k^2_z=\varepsilon\mu\bigl(\frac{\omega}{c}
\bigr)^2$, without making at least one component of the wave
vector imaginary. Such a medium cannot support propagating waves
at all and the wave amplitudes always decay exponentially with
distance inside the medium. If the dielectric constant is a
purely real and negative number, then the Poynting vector is
identically zero, corresponding to a dissipationless metal. Purely
evanescent waves do not transmit energy, and so cannot
give rise to dissipation.

Things become a little more complicated if the medium supporting
the evanescent wave has a dielectric constant which has not only a
real part that is negative, but also a small imaginary part
corresponding to absorption in the medium. In this case the
magnetic field associated with the wave in the medium takes
the form (assuming that the electric field of the incident
radiation lies in the plane of incidence),
\begin{equation}
{\bf {\cal H}}({\bf r},t)=[0~1~0] {\cal H}_0\,
e^{ik_xx-{k}_z'z-i{k}_z''z-i\omega t},
\end{equation}
where [0~1~0] denotes a unit vector in the y-direction. Because we
are dealing with an evanescent mode with
$k_x^2>\bigl(\frac{\omega}{c}\bigr)^2{\rm{Re}}[{\varepsilon\mu}]$,
we have made the transformation $k_z\rightarrow ik_z$ with real
and imaginary parts ${k}_z'$ and ${k}_z''$, respectively. The
electric field can be obtained from Maxwell's equation, $c {\bf
k}\times{\bf {\cal H}}= -\omega e{\bf {\cal E}}$. Although the
imaginary part of the dielectric constant does not affect the
amplitude of the evanescent wave, it induces a phase shift as a
function of the distance into the medium (that is, the
$z$-direction). The Poynting vector is now finite and given by
\begin{equation}
\langle {\bf S}\rangle_t= \Bigl[ \varepsilon_r k_x,~0, ~
\varepsilon_r {k}''_z+\varepsilon_i {k}'_{z}\Bigr] \frac{c^2
|{\cal H}_0|^2}{8\pi\omega \vert \varepsilon \vert^2}
e^{-2k'_zz}, \label{pv}
\end{equation}
where $\varepsilon=\varepsilon_r+i\varepsilon_i$.
Because $k_z=\sqrt{k_x^2-\varepsilon\mu\bigl(\frac{\omega}{c}\bigr)^2}$,
we see that
\begin{equation}
{k}''_z\simeq-\frac{\mu\varepsilon_i(\omega/c)^2}{2\sqrt{k_x^2-
\varepsilon_r\mu(\omega/c)^2}},
\end{equation}
that is, ${k}''_z\propto \varepsilon_i$. In other words, the Poynting
vector is only finite if $\varepsilon_i\ne 0$. Hence the energy
dissipation ($\sim -d S/dz$) arises, as before, from the imaginary
part of the dielectric constant although the exponential decay of
the Poynting vector is due to the real part.

The roles of the real and imaginary parts of the dielectric
constant in affecting the amplitude of an evanescent wave are the
opposite of those for a propagating wave. The decay in the
amplitude of an evanescent wave is due to the real (and perhaps,
negative) part of the dielectric constant. In contrast, the
imaginary part of the dielectric constant, to first order in
$\varepsilon_i$, only causes a phase shift with respect to the
distance into the medium. This phase shift can be
measured experimentally under suitable conditions.\cite{sheng} The
phase shift as a function of position is crucial: it means that
the Poynting vector of the evanescent wave is finite and energy is
absorbed by the medium.

\section{The case of electrons}
Matter waves, like electrons, are described by a wave function
which, for non-relativistic energies, satisfies the unitary
Schr\"{o}dinger equation. The unitarity of Schr\"{o}dinger's
equation implies that probability is preserved. This condition
reflects the fact that massive particles, in contrast to photons,
cannot be physically absorbed. However, in mesoscopic systems
there are many situations in which electrons in one coherent
channel are scattered inelastically due to electron-electron and
electron-phonon interactions into other channels.\cite{datta} Thus
when the transmission through the coherent channel is measured
using an interferometer, it appears as if the electrons in the
specified channel are absorbed. To describe this problem
completely, a substantial generalization of the usual scattering
formalism would be required to include all relevant interactions
and processes. However, the problem can be simplified dramatically
by modeling the apparent absorption of electrons using a
phenomenological approach based on an imaginary
potential.\cite{stonelee,jonson,zohta}

Imaginary potentials are often introduced in the context of scattering
theory and the optical theorem.\cite{schiff} The presence of an imaginary
potential destroys the unitarity of Schr\"{o}dinger's equation. For a
complex potential of the form $V({\bf r})=V_r({\bf r})-iV_i({\bf r})$
(with $V_i({\bf r})$ a positive definite function), the equation of
motion for a wavepacket $\Psi({\bf r},t)$ is
\begin{equation}
i\hbar\frac{\partial \Psi({\bf r},t)}{\partial t}=-\frac{\hbar^2}{2m}\nabla^2
\Psi({\bf r},t)+\bigl[V_r({\bf r})-iV_i({\bf r})\bigr]\Psi({\bf r},t).
\end{equation}
 The importance of the minus
sign in the potential becomes apparent from the equation of motion for the probability
density,
$P({\bf r},t)=\Psi^*({\bf r},t)\Psi({\bf r},t)$,
\begin{equation}
\frac{\partial P({\bf r},t)}{\partial t}+\nabla \cdot {\bf S}({\bf
r},t)=-\frac{2V_i}{\hbar} P({\bf r},t),
\end{equation}
where ${\bf S}$ is the particle flux. The imaginary potential clearly acts as a
sink for particles.

The imaginary potential in the Schr\"{o}dinger equation can be
eliminated by writing the wavefunction as $\Psi({\bf
r},t)=\exp(-V_i t/\hbar)\phi({\bf r},t)$, where $\phi({\bf
r},t)$ obeys the Schr\"{o}dinger equation in the absence of the
imaginary potential. Thus an imaginary potential in the
Schr\"{o}dinger equation is equivalent to a wavefunction that
intrinsically decays in time. Thus it is not surprising that
imaginary potentials have been widely used to construct
phenomenological models for absorption in quantum mechanics.

If the imaginary potential is spatially uniform, then in the
absence of any sources of particles, the decay of probability will
be uniform in time. In this case it makes sense to examine the
steady-state scattering solutions, that is, solutions of the form
$\Psi({\bf r},t)=\psi({\bf r}) \exp(-iEt)$, which obey (in
one-dimension) the time-independent wave equation
\begin{equation}
-\frac{\hbar^2}{2m}\frac{d^2\psi(x)}{dx^2}+(V_r-iV_i)\psi(x)=E\psi(x),
\end{equation}
where $E$ is the energy of the wavefunction. For constant
potentials and $E>V_r$, $\psi(x)$, represents propagating
waves of the form $\psi\sim \exp(ikx)=\exp(ik'x-k''x)$,
where
\begin{equation}
k=k'+ik''\simeq\sqrt{2m(E-V_r)/\hbar^2}\Bigl[1+i\frac{V_i}{2(E-V_r)}\Bigr],
\end{equation}
assuming $V_i\ll E-V_r$. The amplitude of the wave decays with
distance because of the imaginary component in the potential, in
close analogy with the optical case where the imaginary part of
the dielectric constant causes a wave to decay in amplitude.

In the classically forbidden region, where $E<V_r$, the
wavefunction corresponds to tunneling rather than propagation,
$\psi = \psi_0 \exp(-Kx)=\psi_0 \exp[-(K'-iK'')x]$,
where
\begin{equation}
K=K'-iK''=\sqrt{2m(V_r-E)/\hbar^2}\biggl[1-i\frac{V_i}{2(V_r-E)}\biggr].
\label{eqbigK}
\end{equation}
This behavior is like the optical case of evanescent waves: the
decay of the wavefunction with distance is controlled by the real
part of the potential, while the imaginary part of the potential
gives rise to just a change in the phase. However, the phase
change is again important because it is responsible for absorption.

\begin{figure}[h]
\epsfig{file=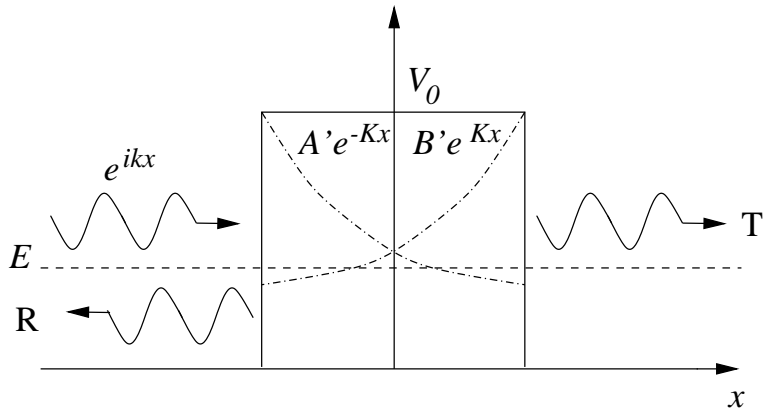, width=10.0cm} \caption{\label{fig2}
Tunneling of a wave across a potential barrier. The dash-dotted
lines inside the barrier schematically show the decaying and
amplifying evanescent waves inside.}
\end{figure}

The absorption that arises from evanescent waves that undergo a
position-dependent phase shift is nicely illustrated by
considering the text-book example of tunneling through a
rectangular barrier,\cite{schiff} shown schematically in Fig.~2.
If the wave has energy $E>V_0$ so that it is propagating
throughout, then the wavefunction inside the barrier region is
given by $\psi = A e^{ikx} + B e^{-ikx}$, where
$k=\sqrt{2m(E-V_0)/\hbar^2}$ and the coefficients $A$ and $B$ are
determined by the boundary conditions for the wavefunction. The
particle flux takes the form
\begin{equation}
J=-\frac{i\hbar}{2m}\biggl[\psi^*\frac{\partial \psi}{\partial x}-
\psi\frac{\partial \psi^*}{\partial x}\biggr]=\frac{\hbar
k}{m}\bigl(|A|^2-|B|^2\bigr).
\end{equation}
That is, for propagating waves the flux is given by the
difference in amplitudes between the forward and backward
travelling wave solutions.

For a wave of energy less than the barrier height, $0<E<V_0$, the
wavefunction inside the barrier has the form,
\begin{equation}
\psi(x)=A' e^{-K x}+B' e^{K x},
\end{equation}
with $K=\sqrt{2m(V_0-E)/\hbar^2}$, and consists of two parts that
decay and amplify with $x$, respectively. In this case, the
particle current is given by,
\begin{equation}
J=-\frac{i\hbar K}{m} ( A'^*B'-B'^*A').
\end{equation}
Crucially, the particle current depends on the phase difference
between $A'$ and $B'$, that is, between the decaying and growing
solutions. For a barrier of finite width with $0<E<V_0$, the
coefficients $A'$ and $B'$ are complex and there is a finite
particle current.

For an incident evanescent wave, the energy is less than zero so
the coefficients $A'$ and $B'$ are real as long as $V_0$ is real.
Hence there is no particle flux in the entire system
--- reaffirming that an evanescent wave does not give
rise to a particle current. However, in the case of an evanescent
wave in an absorbing medium with $\psi = \psi_0 \exp [ -Kx]$
$K$ given by Eq.~(\ref{eqbigK}), the particle current is
\begin{equation}
J = \frac{\hbar K''}{m}\vert \psi_0 \vert^2 \exp(-2K' x).
\end{equation}
Here the change of phase due to the imaginary part of the
potential causes a finite amount of absorption, and there is a
finite particle current corresponding to absorption in the medium.

\section{Imaginary potentials and scattering}
As we have seen, an imaginary potential in Schr\"{o}dinger's
equation can account phenomenologically for absorption in a flux
of electrons. This approach has been applied to a number of
problems in mesoscopics such as the effect of inelastic processes
on tunneling resonances in one dimension.\cite{stonelee} However,
as Rubio and Kumar\cite{rk} have pointed out, an imaginary potential causes scattering
in addition to absorption and this means that it must be used with great care. The
scattering due to an imaginary potential introduced to model inelastic processes is
frequently neglected (see Refs.~\onlinecite{rk}, \onlinecite{stonelee}, and
\onlinecite{jonson}), but such an approximation is hardly warranted because there is no
guarantee that the scattering should be weak.

A localized imaginary potential causes scattering, just like a
real potential, because of the inevitable mismatch in the overall
potential where the imaginary potential begins and ends. We can
see this effect more clearly by considering the example used in
Ref.~\onlinecite{rk} of a complex scatterer located at the origin,
$V(x)=(V_r-iV_i)\delta(x)$. We use a delta-function potential
because of its simplicity, even though it doesn't resemble the
potential in any realistic mesoscopic system. However, the main
features of this simple example carry over to much more realistic
mesoscopic systems such as double barrier potentials.

\begin{figure}[h]
\includegraphics[height=200pt,width=200pt]{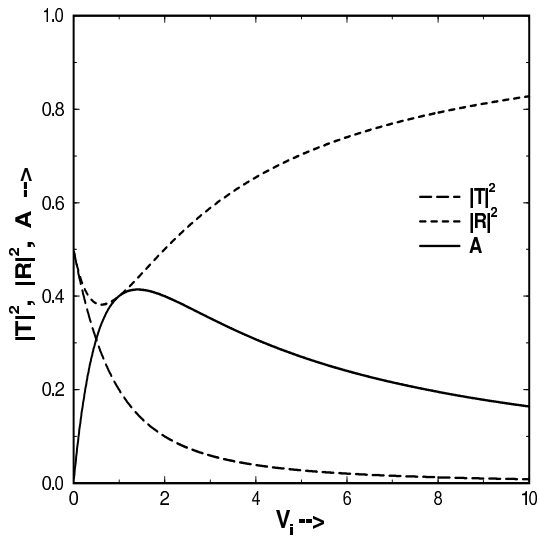}
\caption{\label{fig3} The absorption (A), reflection ($\vert R
\vert^2$), and transmission ($\vert T \vert^2$) coefficients for a
wave incident on a $\delta$ potential of strength $V_r -iV_i$. The
real part $V_r = 1$ and $m/(\hbar^2 k) = 1$ in the graph.}
\end{figure}

By solving the Schr\"{o}dinger equation for a plane wave,
$\psi(x)\propto e^{ikx}$, incident from the left
($x=-\infty$), we can obtain the transmission ($T$) and reflection
($R$) coefficients associated with the potential,\cite{rk}
\begin{equation}
|T|^2=\biggl |\frac{1}{1-\xi(V_r-iV_i)} \biggr |^2, \label{T}
\end{equation}
\begin{equation}
|R|^2=\biggl |\frac{\xi(V_r-iV_i)}{1-\xi(V_r-iV_i)} \biggr |^2,
\label{R}
\end{equation}
where $\xi=m/(i\hbar^2k)$. The absorption due to the potential,
$A=1-|T|^2-|R|^2$, is given by
\begin{equation}
A=\frac{2mV_i/\hbar^2k}{(1+mV_i/\hbar^2k)^2+(mV_r/\hbar^2k)^2}. \label{A}
\end{equation}
Notice that the absorption goes to zero when $V_i=0$.
In contrast, $R\neq 0$ when $V_r$ is zero, but $V_i$ is finite.
Indeed, the behavior of $R$, $T$, and $A$ with increasing $V_i$ is
surprisingly complex, as shown in Fig.~3. Crucially, the
absorption does not increase steadily with $V_i$, but instead goes
through a maximum and then decreases as $V_i$ is
increased further. Thus the idea that the magnitude of the
imaginary potential is a simple measure of absorption is
incorrect. These surprising properties are by no means peculiar to the
$\delta$-function potential, but apply
quite generally to complex potentials.\cite{rk,jayan,sar_prb} We
should point out, however, that it is possible to engineer some
complex (real + imaginary) spatially varying potentials that
completely absorb a wave without reflection or transmission at
selected wave numbers.\cite{muga} Carefully engineered potentials
not withstanding, the basic lesson remains: it is not possible
to model inelastic processes with complete fidelity in mesoscopic
systems by simply adding an imaginary part to the potential.
Such an ansatz always gives rise to
additional scattering, which is an artifact of the model rather than
physically real.

For light, an imaginary component in the dielectric constant
generally gives rise to scattering as well as absorption. However,
unlike the case of electrons, the scattering is not an artifact of
an oversimplified model of absorption, but is physically real. For
electromagnetic systems we are interested in learning how we
should tailor the properties of a material in order to observe
certain types of behavior. In contrast, in the case of electrons
we want to see how to tune the potential in order to faithfully
model the physics of a particular system.

In electromagnetism there are exceptions to the link between
absorption and scattering. In other words, we can conceive of a
material in which there is absorption but no reflection. Consider,
for example, the case of light passing from air to a medium with
$\varepsilon=\mu=1+i\Delta$ at normal incidence. The reflection
amplitude at the interface between the media is $R=(Z-1)/(Z+1)$
where $Z=\sqrt{\mu/\varepsilon}$ is the impedance. Clearly there
will be no reflection for $\varepsilon=\mu$ as $Z=1$. However,
there will still be absorption because the refractive index,
$n=\sqrt{\varepsilon\mu}=1+i\Delta$, is complex. The same effect
can be obtained by making the imaginary parts of both
$\varepsilon$ and $\mu$ very large compared to the respective real
parts. This effect turns out to be the basis of the ``stealth''
materials: a material that absorbs electromagnetic radiation
without reflection sends no signal back to a radar and so, in
principle, is undetectable. The breaking of the link between
absorption and reflection for electromagnetic waves is possible
because there are more degrees of freedom to play with compared to
the case of electrons. In the Schr\"{o}dinger equation we are only
able to make the potential complex, but in optical materials both
the dielectric constant and the permeability constant can be made
complex leading to a richer range of phenomena.

In passing, we note that optical media are amplifying when
$\varepsilon_i<0$. In contrast to electrons, coherent amplification of light is
well-known and arises from stimulated emission --- a characteristic unique to bosons.
A mismatch in the imaginary part of the dielectric constant alone can cause resonant
enhancement of the scattering coefficients when the scatterer is amplifying. For
electrons, a positive imaginary component of the potential would formally correspond
to the amplification of the particle flux. Indeed, if the potential is a purely
imaginary
$\delta$-function, $V=-iV_i\delta(x)$, then the reflection and transmission
coefficients diverge when $V_i=-\hbar^2k/m$, as is clear from Eqs.~(\ref{T}) and
(\ref{R}). However, there exists no analogous process for electrons that can give rise
to coherent amplification of the particle flux.

\section{Perfect absorbers}

In practice, maximizing the absorption is surprisingly
complicated: increasing the imaginary part of the potential (for
electrons) or the imaginary part of the dielectric constant (for
light) does not simply equate to increased absorption; rather the
effect of absorption can be drowned out by scattering. In optics,
the way to produce absorption without reflection is to exploit the
magnetic properties of the medium. However, for electrons there is
no easy way of eliminating reflection entirely,\cite{muga} so the
question becomes one of optimizing the amount of absorption
relative to the amount of reflection.

The problem of maximizing the absorption for the canonical case of
electrons incident on a (real) double barrier potential with a
constant imaginary potential in the well between the barriers was
considered in Ref.~\onlinecite{rk}. In this case electrons that
have energies close to those of the levels inside the well undergo
resonant transmission. In the absence of an imaginary potential
the electrons spend a time $\sim\hbar/\Gamma_e$ in the
well,\cite{datta} where $\Gamma_e$, the elastic resonance width,
is determined by the size of the (real) potential barriers.  The
trick for maximizing absorption is to ensure that the electrons
are trapped in the potential well for as long as possible, while
also making $V_i$ (which gives rise to an inelastic resonance
width $\Gamma_i\propto V_i$) as large as possible so that the
absorber can gradually ``bleed'' the probability. However, making
the imaginary part of the potential too large leads to additional
scattering which starts to destroy the resonance and thus reduce
the time spent by the electron in the well. It turns out that the
best compromise between dwell-time and absorption is achieved when
the elastic and inelastic widths of the resonance are matched,
$\Gamma_e\simeq \Gamma_i$.

For non-magnetic optical media, maximizing the amount of
absorption is also non-trivial. One particularly interesting case
is that of surface plasmon modes at an air-metal interface where
the analogy between optical and electronic systems can be used to
devise a way of maximizing the absorption of light by making use
of a plasmon resonance. At a resonant frequency, $\omega$, the
real part of the metal's dielectric constant,
$\varepsilon_m'(\omega) = -\varepsilon_a$, where $\varepsilon_a$
is the dielectric constant of air, and surface plasmon modes on
the metal-air interface can be excited.\cite{raether} These are
purely longitudinal solutions of Maxwell's equations and
correspond to charge density oscillations on the metallic
surface. A metal's dielectric permittivity disperses with
frequency and the resonance condition can be achieved at some
other frequency by swapping air for another medium with a suitable
dielectric constant. The surface plasmon modes can be thought of
as bound states at the metal surface because their amplitude decays
exponentially on both sides of the air-metal
interface. However, for a dissipative metal the plasmons cannot
exist as stationary states.

Under normal circumstances, the surface plasmon modes cannot be
directly excited on a smooth surface by an incident propagating
electromagnetic wave, because the wave vector of the surface plasmon is
larger than the photon wave vector ($|{\bf k}_{\rm sp}|>
\omega/c$).\cite{fn2} However, it is possible to excite the
surface modes on a structured metal surface (grating) or a rough
metal surface by a propagating wave. For a metal film on a
dielectric medium, we can again excite surface plasmons on the
air-metal surface by impinging light at certain angles from inside
the dielectric medium. It is also possible to resonantly excite
the surface plasmon modes by an incident evanescent wave arising,
for example, from the near field modes of an object in air.

Let us consider the case of an evanescent mode incident on a metal
surface from air, with dielectric constants
$\varepsilon_m=\varepsilon_m'+i\varepsilon_m''$ and
$\varepsilon_a$, respectively. For $p$-polarized light (the electric
field vector lies in the plane of incidence), the magnetic field
on the air side of the interface is given by
\begin{equation}
{\bf {\cal H}}=[0~1~0] {\cal H}_0 e^{i(k_x x-\omega t)}
\Bigl[ e^{-k^{(a)}_zz} +R e^{k^{(a)}_z z}\Bigr],
\end{equation}
where,
\begin{equation}
R=\frac{(k_z^{(a)}/\varepsilon_a-k_z^{(m)}/\varepsilon_m)}
{(k_z^{(a)}/\varepsilon_a+k_z^{(m)}/\varepsilon_m)},
\end{equation}
is the Fresnel coefficient for reflection, which is obtained by
matching the tangential components of the electric and magnetic
fields at the interface.\cite{bornandwolf} Similarly, the magnetic
field on the metal side of the interface is given by,
\begin{equation}
{\bf {\cal H}}=[0~1~0] {\cal H}_0 e^{i(k_x x-\omega t)}
\Bigl[ T e^{-k^{(m)}_zz} \Bigr],
\end{equation}
where
\begin{equation}
T=\frac{2k^{(m)}_zk^{(a)}_z/\varepsilon_m\varepsilon_a}{\Bigl(k^{(m)}_z/\varepsilon_m
+k^{(a)}_z/\varepsilon_a\Bigl)}
\end{equation}
is the Fresnel coefficient for transmission,
and $k_x^2-(k_z^{(m)})^2=\varepsilon_m\bigl(\frac{\omega}{c}\bigr)^2$.

If we calculate the Poynting vector on the metal side, we obtain an
expression very similar to that in Eq.~(\ref{pv})
\begin{equation}
\langle {\bf S}\rangle_t= [\varepsilon_m' k_x,~0,~\varepsilon_m'
k''_z+\varepsilon_m'' k'_{z}] \frac{c^2}{8\pi
\omega \vert \varepsilon_m \vert^2}|T|^2 {\mathcal{H}}_0^2 e^{-2k_z'z},
\end{equation}
where $k_z''={\rm Im}[k_z^{(m)}]$ and $k_z'={\rm Re}[k_z^{(m)}]$.
Although the energy flow is predominantly along the surface for
large $k_x$, there is a component normal to the boundary that
corresponds to the energy dissipated in the metal.

For large $k_x$, $|T|^2\sim 1/(\varepsilon''_m)^2$ at the
resonance (when $\varepsilon'_m=-1$). Under these
conditions the absorption is maximized in the limit that
the imaginary part of the dielectric constant goes to zero.
Increasing the value of $\varepsilon_m''$ causes a decrease in
absorption because the consequent mismatch in dielectric constants
at the boundary causes reflection and prevents the wave from even
entering the metal.

The conditions for maximizing absorption in this optical system
turn out to be very similar to those for electrons traversing a
double potential well. In either case absorption arises from a
small imaginary part to the dielectric constant or potential in
combination with a resonance due to the corresponding real part.
In both cases, the resonance keeps the wave in the vicinity of the
absorber long enough for it to suck energy (density) out of the
wave.

\section{Conclusions}

We have discussed aspects of light and electron
wave motion in dissipative media. Specifically, we have explored
the problems associated with propagating and evanescent light in
absorbing media and by analogy the inelastic scattering of
electrons in mesoscopic systems. Adding an imaginary part to the
dielectric constant for light, or the potential for electrons,
can be used to model absorption, but it also causes additional
scattering. This scattering
complicates the use of such phenomenological models to describe
inelastic processes for electrons because the additional scattering is
an unphysical artifact of the model. In fact,
the spurious scattering due to the imaginary part can dominate and actually
prevent absorption of the particle flux if it is large enough,
which would clearly invalidate the model completely.

Finally, we discussed the problem of maximizing the absorption of
a wave. The existence of a bound state for electrons greatly aids
the absorption as it ensures that the electron remains in the
region of absorption for a long time while a small amount of
absorption slowly bleeds the wave without scattering it out of the
bound state.  In this case the analogy between electrons and light
can be used to suggest a way of maximizing absorption of light in
a non-magnetic medium by exploiting a resonant surface plasmon
state.

\begin{acknowledgments}
SAR would like to thank N. Kumar and J. B. Pendry for very
enlightening discussions and comments.
\end{acknowledgments}


\begin{thebibliography}{99}
\bibitem{foot}Usually the absorption in a material varies with the frequency and is
accompanied by dispersion in the real part of the material
parameters as well. In fact, the real and imaginary parts of the
material parameters are related by the well-known Kramers-Kronig
relations. In this paper, however, we only consider a single
well-chosen frequency/energy, which is a good approximation for the
case when the frequency bandwidth is very small.
\bibitem{fn1} For magnetic media, a similar contribution will arise
from the magnetic polarizability.
\end{thebibliography}
\end{document}